\documentclass[prd,aps,10pt,nopacs,twocolumn,superscriptaddress,notitlepage,nofootinbib,nobibnotes,amssymb,amsmath]{revtex4-1}
\pdfoutput=1
\usepackage{graphicx}% Include figure files
\usepackage{dcolumn}% Align table columns on decimal point
\usepackage{bm}% bold math
\usepackage{multirow}

\newcommand{\beq}{\begin{equation}}
\newcommand{\eeq}{\end{equation}}
\newcommand{\beqa}{\begin{eqnarray}}
\newcommand{\eeqa}{\end{eqnarray}}
\newcommand{\beqn}{\begin{eqnarray}}
\newcommand{\eeqn}{\end{eqnarray}}
\newcommand{\eq}[1]{(\ref{#1})}
\newcommand{\bs}{\boldsymbol}
\begin{document}

\title{Machine-learning physics from unphysics: Finding deconfinement temperature in lattice Yang-Mills theories from outside the scaling window}

\author{D.~L.~Boyda}\email{Part of the work was done at the Center for Theoretical Physics, Massachusetts Institute of Technology, Cambridge, MA 02139, USA. Present address: Argonne National Laboratory, Lemont, IL, 60439, USA}
\affiliation{Pacific Quantum Center, Far Eastern Federal University, 690950 Vladivostok, Russia}

\author{M.~N.~Chernodub}
\affiliation{Pacific Quantum Center, Far Eastern Federal University, 690950 Vladivostok, Russia}
\affiliation{Institut Denis Poisson CNRS/UMR 7013, Universit\'e de Tours, 37200 France}

\author{N.~V.~Gerasimeniuk}
\affiliation{Pacific Quantum Center, Far Eastern Federal University, 690950 Vladivostok, Russia}

\author{V.~A.~Goy}
\affiliation{Pacific Quantum Center, Far Eastern Federal University, 690950 Vladivostok, Russia}

\author{S.~D.~Liubimov}
\affiliation{Pacific Quantum Center, Far Eastern Federal University, 690950 Vladivostok, Russia}

\author{A.~V.~Molochkov}\email{molochkov.alexander@gmail.com}
\affiliation{Pacific Quantum Center, Far Eastern Federal University, 690950 Vladivostok, Russia}

\date{\today}% It is always \today, today,
             %  but any date may be explicitly specified

\begin{abstract}
We study the machine learning techniques applied to the lattice gauge theory's critical behavior, particularly to the confinement/deconfinement phase transition in the SU(2) and SU(3) gauge theories. We find that the neural network, trained on lattice configurations of gauge fields at an unphysical value of the lattice parameters as an input, builds up a gauge-invariant function, and finds correlations with the target observable that is valid in the physical region of the parameter space. In particular, if the algorithm aimed to predict the Polyakov loop as the deconfining order parameter, it builds a trace of the gauge group matrices along a closed loop in the time direction. As a result, the neural network, trained at one unphysical value of the lattice coupling $\beta$ predicts the order parameter in the whole region of the $\beta$ values with good precision. We thus demonstrate that the machine learning techniques may be used as a numerical analog of the analytical continuation from easily accessible but physically uninteresting regions of the coupling space to the interesting but potentially not accessible regions.
\end{abstract}

                             % Classification Scheme.

\maketitle

\section{Introduction}

The theory of strong interactions, Quantum Chromodynamics (QCD), exhibits several nonperturbative properties that lack so far a solid theoretical explanation. This theory challenges scientists with the phenomena of confinement of color, mass-gap generation, and chiral symmetry breaking observed at low temperatures. At high enough temperature, QCD experiences a smooth deconfinement transition of the crossover type, where these properties are gradually lost, leaving the scene for various thermal effects. High-temperature properties of QCD matter can be probed experimentally in relativistic heavy-ion collisions that create a quark-gluon plasma that once existed in the early moments of our Universe~\cite{Bzdak:2019pkr}.

The nonperturbative physics of QCD appears as a result of the gluon dynamics encoded in the non-Abelian gauge sector of the theory. Theoretically, these issues can be addressed either in low-energy effective models or in first-principle numerical simulations in a lattice formulation of the theory. In practice, however, the quark matter at finite baryon density poses a substantial challenge even for first-principle numerical simulations due to the notorious sign-problem~\cite{deForcrand:2010ys}. While particular methods, such as analytical continuation, can solve this problem for a low-density plasma, the most advanced lattice QCD approaches encounter difficulties in dealing with the moderate-density quark matter~\cite{DElia:2018fjp}. The dense regime is particularly interesting as it will emerge in the next-generation experiments to be performed at the NICA facility in Dubna, Russia and FAIR facility at Darmstadt, Germany.

One of the promising ways to engage the unsolvable problems in lattice field theories, such as QCD, is based on the application of the newest information processing methods in combination with standard Monte-Carlo techniques. In this work, we aim to discuss machine learning (ML) approaches~\cite{ref:book-ml:1} in the context of non-Abelian gauge theories. We focus on a pure Yang-Mills theory without fermions in order to elucidate the finite-temperature deconfinement phase transition, with a further intention for future applications to the non-Abelian theory with fermions.

It is widely believed that the ML approaches may prove their usefulness in revealing complex mechanisms of nonperturbative phenomena in systems with many (or even infinite, in the thermodynamic limit) degrees of freedom~\cite{ref:review:1,ref:review:2}. The neural network may learn a physical phenomenon in an extensive system by building an approximation of some input parameters' function and mapping it to the target physical observable. This procedure may give an insight into the physical mechanism of the original phenomenon in question by analyzing the way what the neural network learned it (see, for example, Ref.~\cite{SU2_ANN}). Field configurations of the quantum field theory and spin systems, viewed as statistical ensembles in the thermodynamic limit, are well-suited for the application of machine-learning techniques, as it was demonstrated in a number of recent works~\cite{Tanaka:2016rtu,scalar_ML,SU2_ANN,Yoon:2018krb,Bachtis:2020dmf,Matsumoto:2019jia,Chernodub:2020nip,Yau:2020emg}.

In this article, we investigate the ability of a neural network to construct gauge-invariant observables based on the analysis of a limited set of lattice configurations. Of particular interest is the ability of a neural network -- trained on data in a narrow range of parameters, or even at a single isolated value -- to predict observables outside the training range. In particular, we will consider the ability of a network trained at a nonphysical point (which lies outside of a continuum limit of the theory), far from a phase transition, to predict critical behavior of the corresponding order parameter in the scaling region which is related to the continuum limit. 

The structure of our paper is as follows. In Section~\ref{ref:aims} we start with a short overview of the machine-learning approaches to the lattice quantum field theories and describe the aims of the current study in the ML context. In Section~\ref{sec:YM} we describe the subject of interest, Yang-Mills theory on the lattice, highlighting the properties of the training regions of the ML algorithms and determination of the phase transition. The main subject of our paper is described in Section~\ref{sec:ML:SUN} where we use the machine learning methods to built up the order parameter of the lattice Yang-Mills theory using the data outside the scaling window of the theory. The last section is devoted to our conclusions.

\section{Machine learning of lattice fields} 
\label{ref:aims}
 
The use of the ML techniques in analysis of lattice gauge theories may pursue different purposes. 

\textit{Speeding up numerical calculations.} Lattice QCD simulations often require vast computer power and capacious data storage, especially in non-Abelian gauge theories with dynamical fermions. This problem largely limits the lattice volume and reduces the accumulated statistics of numerical simulations. Besides improving the computing hardware, further development of the lattice QCD applications requires radical amelioration of the existing algorithms. The ML methods provide us with exciting options for an advance in this direction.

In this approach, an neural network is trained to recognize certain observables at a limited number of lattice  Monte-Carlo configurations corresponding to a preliminary chosen set of lattice parameters. A well-trained neural network is then supposed to be able to predict the values of these observables for a previously-unseen lattice configuration in a broader range of parameter space. Basically, the machine-learning algorithm works as an improved tool which efficiently makes interpolation and extrapolation in the space of thermalized configurations based on a small number of learned reference points. While the learning phase of the neural networks could be rather slow, a well-rained neural network gives its predictions very fast. The use of the fast neural network instead of the slow Monte-Carlo simulations may significantly reduce the required computing power in computing observables over a wide range of the parameter space. As an example, we mention that this approach shows essential advantages in estimating the constant physics line and the ability to overcome critical slowing down.~\cite{MIT_ml}.

The other direction of improving Lattice QCD simulations is the speeding up of configurations generation and decreasing autocorrelation time. The ML techniques allow generalizing Hamiltonian Monte Carlo Algorithm (state-of-the-art algorithm in Lattice QCD) with neural networks. Authors of~\cite{Levy_HMC} argue considerable improvement in effective sample size and better mixing properties when HMC is stuck in one vacuum. Various approaches, such as Generative adversarial networks~\cite{Urban:2018tqv} or normalizing flows~\cite{Albergo:2019eim}, can be applied to the generation of lattice configuration itself. Latter approach has been recently applied to SU(N)lattice gauge theory~\cite{Boyda_Sampling} and shown considerable improvement of autocorrelation time in U(1) Lattice gauge theory~\cite{Kanwar_Sampling}.

The supervised machine learning was shown to be used as an efficient reweighting technique to extrapolate the Monte-Carlo data over continuous ranges in parameter space~\cite{Bachtis:2020dmf}.

The ML techniques can may be used to uncover the position of a phase transition in the phase space of a model. The key observation is that while the ML algorithm can give robust results at both sides of (and sufficiently far enough from) the phase transition, the neural network becomes less confident as the transition line is approached. This lack of confidence plays a positive role in determination of the phase diagram via ML-based methods. The confusion of the machine-learning algorithm may be quantified via a specific ML variable and may therefore serve as an ML-based order parameter used to determine the location of a phase transition~\cite{ref:phases:3}. This criterion, applied to Abelian monopoles, gives a good prediction of a thermodynamically smooth deconfinement phase transition of the Berezinskii-Kosterlitz-Thouless type in a low-dimensional model that exhibits the confinement phenomenon~\cite{Chernodub:2020nip}.

The ML techniques can also speed up the classification of complicated (nonlocal) observables, for example, of the topological charge in lattice Yang-Mills theory~\cite{Matsumoto:2019jia}.

\textit{Uncovering underlying physics.} The ML methods are instrumental in the exploration of large datasets to reduce complexity and find new features and correlations. These features motivate the application of the ML methods to the high-energy physics experiments to uncover and characterize new particle reactions~\cite{ML_HEP}. The lattice field configurations, generated by the Monte-Carlo techniques rather than by particle experiments, may also be scrutinized by the ML techniques to determine unknown correlations and pinpoint new physics. 

Undoubtedly, the neural network itself can not identify a new mechanism of the studied phenomenon. Instead, the method provides an efficient numerical tool to find new relationships between field observables hidden otherwise in vast volumes of the data (field configurations). The nature of the new relations -- provided by the ML algorithm -- gives information for a researcher to pinpoint the physical mechanism of the phenomenon.

A neural network may uncover necessary ingredients of a mechanism that drives a phase transition. One of the essential tasks of an ML algorithm is the phase classification. The classification is done on the basis of lattice configurations that contain all information about the non-perturbative physics of the modeled system. In the process of learning how to classify the phases, the machine-learning algorithm studies the lattice data that was previously generated by the Monte-Carlo algorithm. The network builds an internal observable (a decision function) that allows it to distinguish the two phases. As soon as the network acquires sufficient skills to distinguish the phases, the constructed decision becomes an object for the further study: it gives explicitly the structure of the variable that the neural network has built to distinguish the phases. A recent discussion of the use of the machine learning techniques for understanding of the phase structure of a lattice field theories through the statistical analysis of Monte Carlo samples may be found in Ref.~\cite{Bluecher:2020kxq}.

The feasibility of this approach has recently been demonstrated in Ref.~\cite{SU2_ANN} where the machine-learning algorithm has constructed -- via a training process -- the phase-sensitive observables in the Ising model and SU(2) Yang-Mills theory on the lattice. It turned out that the decision functions give, respectively, the mean magnetization and the Polyakov loop variable which are indeed well known order parameters that determine the phase structure in these theories.

\textit{Application to problems unreachable with traditional methods.} An essential advantage of the machine learning methods is that they can be applied to certain physical phenomena which cannot be simulated with traditional Monte Carlo methods. For example, the authors of Ref.~\cite{Broecker:2017} demonstrated that convolutional neural networks may identify and locate phase transitions for quantum many-fermion systems that experience severe fermion sign problem where conventional approaches fail. Notice, the ML model did not have any knowledge about the Hamiltonian of the system. This result demonstrates the power of the neural network, and the ability to make physically sound predictions. 

In our paper, we aim to solve yet a different problem with the ML methods. Let's assume that we have a lattice system where the traditional Monte Carlo methods work in a restricted domain of the parameter space that cannot be extrapolated to the continuum limit. The importance of this unphysical and seemingly useless region is that in this particular domain of coupling constants we know the value of the order parameter very well, in a contrast with the physically interesting critical region. we demonstrate the power of the ML algorithm which is able to make correct predictions in the interesting domain of the coupling space after being trained in an unphysical single point of the model.
The prediction requires lattice configurations at the prediction points. Thus,  this approach does not solve the lattice configuration generation's question in problematic areas. It is rather a tool for extrapolating observables to the areas where direct calculations are difficult or impossible.

Our study is motivated by the unsolved status of the QCD phase diagram at nonzero baryonic density, where the results are available at low values of the baryonic chemical potential. The interesting region of the parameter space, that contains a critical endpoint, cannot be reached with the direct Monte Carlo simulations. The moderate-density region is accessible only with a combination of analytical and numerical tools such as Taylor expansion and analytical continuation (see, for example, a recent review in Ref.~\cite{DElia:2018fjp}). In this sense, our machine learning approach may be considered as a purely numerical technique that serves as an analytical continuation tool.

To demonstrate the power of the method, we take a well studied model as a playground. We consider the lattice Yang-Mills theory for two and three colors, train an neural network to guess an order parameter on the configurations with a  small physical volumes in a perturbative regime, and then show that the ML method may extrapolate (``analytically continue'') 
the results to the critical confining-deconfining region, using its lattice configurations as input.

\section{Yang-Mills theory at finite temperature}
\label{sec:YM}

We consider lattice SU(N) gauge theories with $N = 2$ and $N = 3$ colors at finite temperature. The lattice theory is formulated via the Euclidean path integral
\beqn
Z = \int \biggl( \prod_l U_l  \biggr)  e^{ - S[U]},
\eeqn
where the integration over the lattice gauge fields $U_l$ that belong to the SU(N) gauge group.

The Wilson action of the lattice SU(N) Yang-Mills theory,
\beqn
S[U] = \beta \sum_{P} \left(1 - \frac{1}{N}{\mathrm {Re}}\, {\mathrm{Tr}\,} U_P\right).
\label{eq:S}
\eeqn
is formulated in the Euclidean spacetime on the lattice with the volume $N_s^3 \times N_t$. The sum runs over the lattice plaquettes $P = \left\{ x,\mu\nu \right\}$ described by the position of a plaquette corner $x$ and the plane orientation with directions $\mu \neq \nu$. The non-Abelian plaquette field $U_P$ is given by the ordered product of the non-Abelian link fields $U_l$ along the perimeter of the plaquette: $U_P = \prod_{l \in \partial P} U_l$. 

The lattice coupling in the action~\eq{eq:S} is related to the gauge coupling $g$ in the continuum limit:
\beqn
\beta = \frac{2N}{g^2}.
\label{eq:beta}
\eeqn
The continuum limit of the lattice Yang-Mills theory~\eq{eq:S} corresponds to the weak-coupling regime, $\beta \to \infty$. 

The length $N_s$ of the shortest lattice direction determines the temperature $T$ of the system:
\beqn
T = \frac{1}{a N_s},
\label{eq:T}
\eeqn
where $a$ is the physical lattice spacing. The critical temperature of SU(2) and SU(3) gauge theories in the continuum limit are, respectively, as follows~\cite{Fingberg:1992ju,Boyd:1996bx}:
\beqn
T_c^{\textrm{\tiny{SU(2)}}} = 0.69(2) \sigma^{1/2}_{\textrm{\tiny{SU(2)}}}, 
\qquad
T_c^{\textrm{\tiny{SU(3)}}} = 0.629(3) \sigma^{1/2}_{\textrm{\tiny{SU(3)}}},\quad
\label{eq:Tc}
\eeqn
where $\sigma_{\mathrm{SU(N)}}$ denotes the corresponding zero-tempe\-ra\-tu\-re string tensions.
%that set the physical mass scales for all dimensionful observables.

The knowledge of the physical value of the lattice spacing $a$ at a given value of the lattice coupling~$\beta$ allows us to relate dimensionful lattice quantities to their continuum counterparts. For example, the value of temperature~\eq{eq:T} at a critical lattice coupling $\beta_c$ allows us to recover the deconfining temperatures in physical units~\eq{eq:Tc}. The continuum limit of lattice Yang-Mills theory is achieved in a weakly-coupling region ($g \ll 1$ or, equivalently, $\beta \gg 1$) where the dependence of the lattice spacing $a$ on the $SU(N)$ coupling constant $g$ is controlled by the renormalization group equation. For example, in pure SU(2) gauge theory a two--loop calculation gives
\beqn
a_{\textrm{\tiny{SU(2)}}}\left( g^2 \right) = \frac{1}{\Lambda_L} \exp\left\{ - \frac{12 \pi^2}{11 g^2} + \frac{51}{121} \ln \frac{24 \pi^2}{11 g^2} \right\},
\label{eq:a}
\eeqn
where $\Lambda_L \simeq 0.0221 \sqrt{\sigma_{\textrm{\tiny{SU(2)}}}}$ is a fixed mass scale. The coupling constants in the continuum $g$ and on lattice $\beta$ are related to each other via Eq.~\eq{eq:beta}.

The Yang-Mills theories possess the confining phase at low temperature and the deconfinement phase at high temperature. The phases are separated by the thermodynamic phase transition. The phase transition in the simplest two-color ($N = 2$) gluodynamics is of the second order while the theories with the $N \geqslant 3$ colors possess the stronger, first-order phase transition. 

The order parameter of the deconfinement phase transition is the expectation value of the Polyakov loop. In the lattice calculations, it is convenient to identify the bulk Polyakov loop:
\beqn
L = \frac{1}{V} \Bigl\langle \Bigl| \sum_{\bs x } L_{\bs x}\Bigr| \Bigr\rangle,
\label{eq:L}
\eeqn
where the sum goes over all spatial sites $\bs x$ of the lattice and $V = N_s^3$ is the spatial volume. The local Polyakov loop,
\beqn
L_{\bs x} = \frac{1}{N} {\mathrm{Tr}\,}\prod_{t = 0}^{N_t - 1} U_{{\bs x},t;4},
\eeqn
is given by the ordered product of the lattice $U_{x,\mu}$ matrices along the temporal direction $\mu = 4$.

It is also convenient to define the susceptibility of the Polyakov loop using a light abuse of notations:
\beqn
\chi_2 \equiv \langle L^2 \rangle - \langle L \rangle^2 = \Bigl\langle \Bigl| \sum_x L_x\Bigr|^2 \Bigr\rangle 
- 
\Bigl\langle \Bigl| \sum_x L_x\Bigr| \Bigr\rangle^2.
\label{eq:chi2}
\eeqn
In the SU(3) gauge theory, we will also work with the real and imaginary parts of the Polyakov loop, which amounts to substitute $L_{\bs x} \to \mbox{Re} L_{\bs x},\, \mbox{Im} L_{\bs x}$ in Eq.~\eq{eq:L} and below. 

Susceptibility of the Polyakov loop is a good order parameter for the determination of the confinement/deconfinement phase transition  and the critical lattice coupling. In our study, we use rather low statistics (about 100 lattice configurations) for the neural network predictions. Therefore, statistical errors are large and they do not allow us to determine the critical value with acceptable errors using the susceptibility only. In this study, we employed another statistical moment, the Binder cumulant~\cite{Binder}:
\begin{equation}
	C_4 = 1- \frac{\langle L^4 \rangle}{3 \langle  L^2 \rangle^2},
\end{equation}
and determine the critical value $\beta_c$ by fitting data of the Binder moments of the Polyakov loop with the help of the following function:
\beqn
C_4^{\mathrm{fit}}(\beta) = A + B \tanh\frac{\beta- \beta_c}{\delta \beta},
\label{eq:Binder}
\eeqn
where $A$ and $B$ are the fitting parameters that determine the strength of the cumulant, while $\beta_c$ and $\delta \beta$ correspond to the position of the transition and its width, respectively. These critial values will be shown in the figures below. 

In the next section, we describe the machine-learning algorithm which includes the training of the neural network. The training points for SU(2) and SU(3) gauge theories are set at the lattice coupling constants $\beta = 4$ and $\beta = 10$, respectively. Both these points correspond to a deep weak-coupling regime where the gluons reside in a perturbatively deconfining phase for the lattice sizes used. In other words, at these parameters, the physical size of the lattice is so small, that the confining string has no space to develop itself. Since all distances in such a volume are smaller than the confining scale, the Yang-Mills theory resides necessarily in a deconfining state. The perturbative deconfinement regime has obviously a different nature compared to the usual deconfinement phase: the former appears as a result of a finite spatial volume while the latter comes as a consequences of finite-temperature effects in the thermodynamic (infinite-volume) limit. 

In order to quantify the scales of the finite-volume deconfinement, we notice that at the large lattice coupling $\beta = 4$, the lattice spacing of the SU(2) gauge theory is 
$a =  3.4\times 10^{-3} \sigma^{-1/2}$. 
In physical units, $a = 1.6\times 10^{-3}\mbox{\,fm}$, where we adopted the phenomenological value $\sqrt{\sigma} = 440\,\mathrm{MeV} \simeq (0.46 \, \mathrm{fm})^{-1}$ for the string tension. The confinement phase may only be realized at spatial lattice sizes $N_s \gtrsim 300$, for which the lattice size is of the order or bigger than the typical confining distance scale, $a_{\textrm{\tiny{SU(2)}}} N_s \gtrsim \sigma^{-1/2}$. For a typical lattice size used in the numerical simulations, $N_s \sim (\mbox{a few}) \times 10$, the vacuum resides in the finite-volume deconfining phase because the maximal possible inter-quark distances are much smaller than the perturbative scale $r \simeq 0.1\,\mathrm{fm}$. Similar estimations are also valid for the SU(3) lattice theory.

The training points $\beta=4$ and $\beta=10$ do not correspond to physically viable realizations of the continuum SU(2) and SU(3) Yang-Mills theories in their thermodynamic limits. These points are selected to represent an uninteresting unphysical region of the theory at which, however, the explicit calculations may be performed with the help of a Monte Carlo technique. We will show that the information coming from the MC configurations are enough for the ML algorithm to learn about the order parameter and make accurate predictions in the physically relevant scaling window of the lattice Yang-Mills theory.

\section{Restoration of the order parameter with neural networks}
\label{sec:ML:SUN}

In this section, we discuss application of the ML methods to predict an order parameter of the theory with lattice configurations as an input. The study focuses on building of an neural network that can predict observables of the SU(2) and SU(3) theories.

\subsection{SU(2) gauge theory}

To build a machine-learning algorithm that can analyze lattice data of non-Abelian theory, we need to construct a multidimensional dataset from a lattice configuration that is a matrices dataset.  To this end, we use the following vector representation for the SU(2) matrices:
\beqn
U {=} 
\left(
\begin{matrix}
u_{11} & u_{12} \\
u_{21} & u_{22}
\end{matrix}
\right)
{\equiv}
\left(
\begin{matrix}
a_1 + i a_2 & a_3 + i a_4 \\
- a_3 + i a_4 & a_1 - i a_2
\end{matrix}
\right)
{\rightarrow}
\left(
\begin{matrix}
a_1 \\
a_2 \\
a_3 \\
a_4 \\
\end{matrix}
\right),
\qquad
\label{input_par}
\eeqn
where $a_1=\mbox{Re}(u_{11})$, $a_2=\mbox{Im}(u_{11})$, $a_3=\mbox{Re}(u_{12})$, and $a_4=\mbox{Im}(u_{12})$.

After the matrix dimension's flattening, an array with shape $[N_t, N_s, N_s, N_s, Dim, 4]$ represents the lattice configuration. The last dimension corresponds to the matrix element numbering discussed above, and $Dim$ is the direction $\mu$ of the matrix $U_\mu(x)$ at every lattice site [Nt, Ns, Ns, Ns]. We use 3D convolutional layers and reshape lattice configuration as a 4D array (3 dimensions for spatial coordinates and one for channels) due to technical reasons. Since we build a neural network that searches correlations between any two matrices $U_\mu(x)$ and $U_\nu(y)$ at the points $x$ and $y$ closed to each other, we merge the last two dimensions of the array. Other two dimensions could be also merged by cost of locality - array $M[y][x]$ can be presented as an array $M[y* N_y + x]$ . 

The resulting lattice data array has a dimension of $4$. The first dimension corresponds to the numbering of temporal layers of the lattice. The second dimension described by single flattened array of two spatial axis, third dimension of array corresponds to the last axis of the spatial direction, and the last dimension corresponds to the numbering of the matrix elements~\eq{input_par} for all lattice directions $\mu$. 

For the lattices with $N_t=2$, the neural network consists of one three-dimensional convolutional layer with 16 filters and the kernel size $2 \times 1 \times 1$ with Relu activation function, and a final dense layer with a linear activation function with 16 neurons. The averaging layer over the entire volume and the flattening layer separate the convolutional and dense layers. The architecture for the temporal lattice extension $N_t = 2$ is shown in Table~\ref{tab:ANN1}.
\begin{table}[htb]
	\begin{tabular}{|c|c|c|}		
		\hline
		Layer & 	\multicolumn{2}{|c|}{Structure} \\
		\hline
		\multicolumn{3}{|c|}{}\\
		\hline \multirow{2}{*}{InputLayer} & In & ($N_t=2$, $N_s {\times} N_s$, $N_s$, $\mbox{Dim} {\times} \mbox{U}$) \\
	   \cline{2-3}& Out &($N_t =2$, $N_s {\times} N_s$, $N_s$, $\mbox{Dim} {\times} \mbox{U}$) \\
		\hline
			\multicolumn{3}{|c|}{}\\
		\hline
\multirow{2}{*}{Conv3D} & In &	($2$, $N_s {\times} N_s$, $N_s$, $\mbox{Dim} {\times} \mbox{U}$) \\
\cline{2-3} & Out & ($1$, $N_s {\times} N_s$, $N_s$, $16$) \\
		\hline
			\multicolumn{3}{|c|}{}\\
		\hline
		\multirow{2}{*}{AveragePooling3D} &In & ($1$, $N_s {\times} N_s$, $N_s$, $16$) \\
\cline{2-3}	 & Out & ($1$, $1$, $1$, $16$) \\
		\hline
			\multicolumn{3}{|c|}{}\\
		\hline
		\multirow{2}{*}{Flatten}& In & ($1$, $1$, $1$, $16$) \\
\cline{2-3}		&Out& ($16$)\\
		\hline
			\multicolumn{3}{|c|}{}\\
		\hline
		\multirow{2}{*}{Dense} &In& ($16$) \\
\cline{2-3}	& Out	  & ($1$) \\
		\hline  
\end{tabular}
	\caption{Architecture of the neural network for the prediction of the Polyakov Loop in the SU(N) gauge theory with the temporal size of the lattice $N_t=2$. Here Dim is dimension of theory, U is dimension of vector representation}.
	\label{tab:ANN1}
\end{table}

It is important to stress that the convolution kernels shape defines the physical observable that the neural network can extract from the lattice data. For example, the kernel size equal to $N_t \times 1 \times 1$ leads to the neural network output with a function of $N_t$ $U_\mu(x)$ matrices located along the closed line in $N_t$ direction that corresponds to the Polyakov loop. 

We generate 9000 lattice configurations at the one value ($\beta=4$) of the lattice coupling for lattices with the spatial sizes $N_s = 8, 16, 32$ and the temporal sizes $N_t = 2, 4$. We also generate 100 configurations for a number of points at lower values of the coupling $\beta$, that the neural network does not use for training but rather for prediction.

Although a study of confinement-deconfinement phase transition does not require configurations from all possible vacuum sectors, we found it essential to have high-quality data generated from different vacuum sectors to train a neural network.

We train the neural network on the lattice configurations generated in the (volume-induced) deconfinement phase at the point $\beta = 4$ for SU(2) that is far from the phase transition point. The neural network is trained to predict correctly the value of the Polyakov loop that is already known from the Monte Carlo simulations. We use the mean squared error (MSE) as a loss function and the Adam algorithm as the neural network parameters' optimization method. The training is done in batches of size 10 - 50 configurations for SU(3) and 10 - 50 for SU(2). The training is halted when the loss function reached a plateau so that the neural network gained the maximal possible -- for the given architecture -- knowledge how to reconstruct the order parameter from the lattice configurations.

\begin{figure*}[htb]
\begin{tabular}{ccc}
\includegraphics[width=0.3\linewidth]{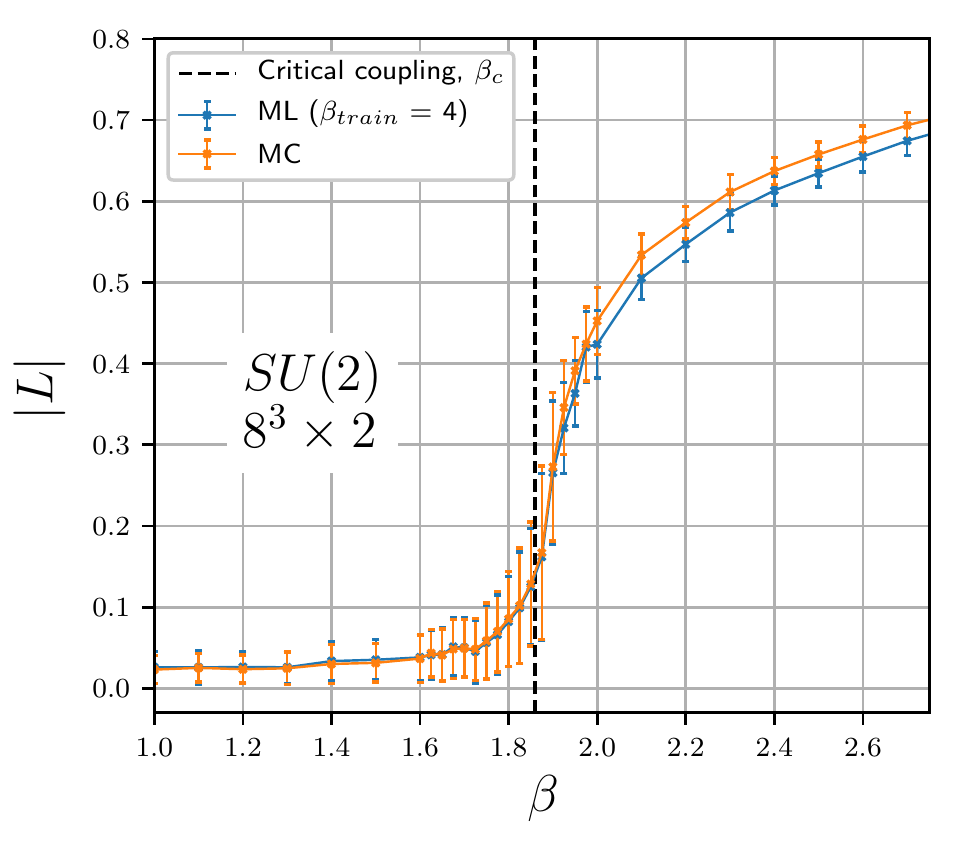} &
\includegraphics[width=0.3\linewidth]{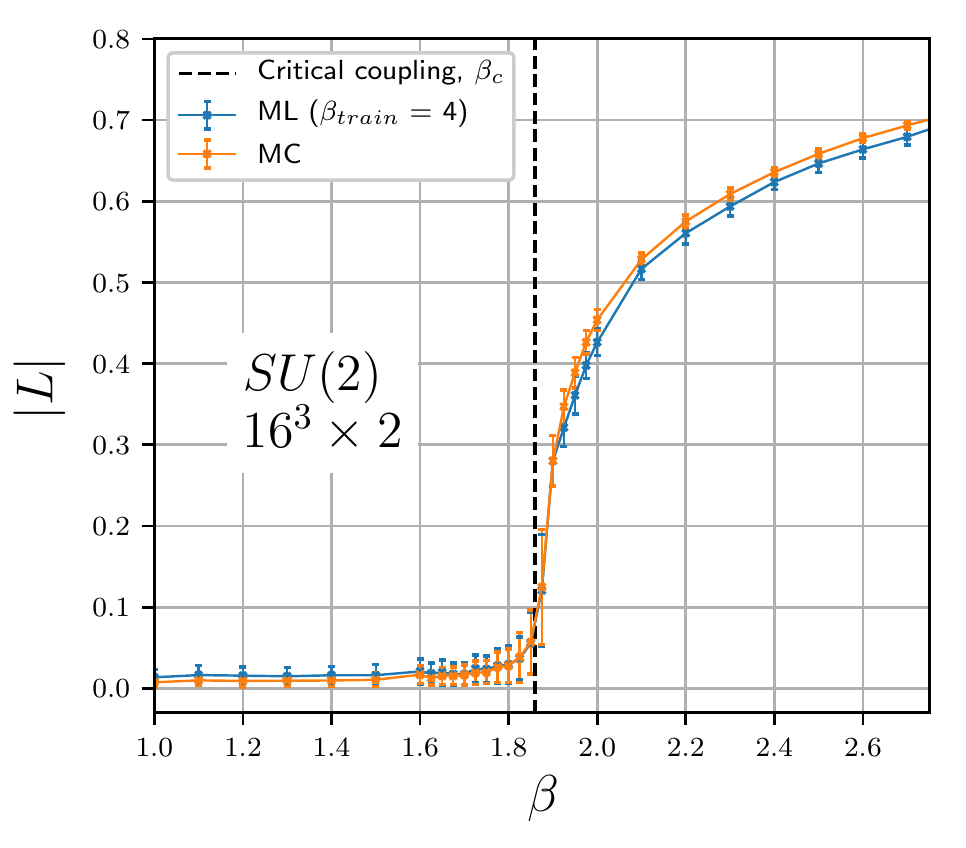} &
\includegraphics[width=0.3\linewidth]{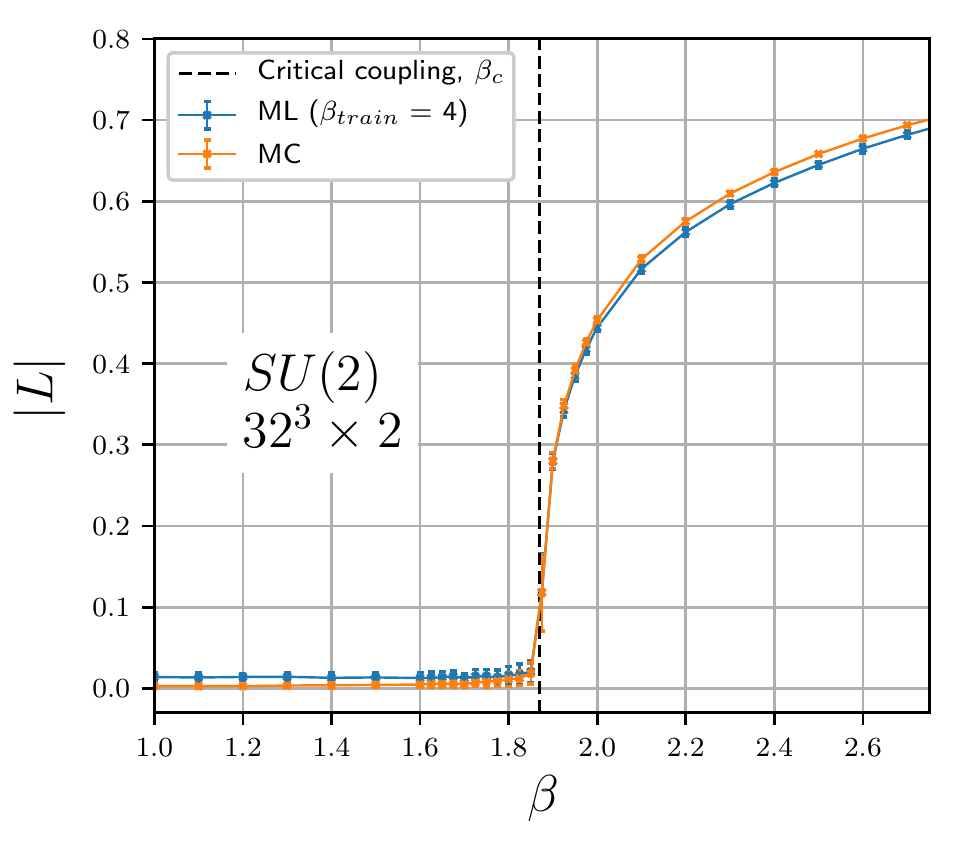}\\[-2mm]
%\hskip 3mm $8^3\times 2$  &
%\hskip 3mm $16^3\times 2$  &
%\hskip 3mm $32^3\times 2$  
\end{tabular}
\caption{The Polyakov loop in SU(2) gauge theory at the $N_t=2$ and $N_s=8,16,32$ lattices. The Monte-Carlo (MC) simulation, shown by the blue line, and the prediction of the machine-learning (ML) algorithm, shown by the orange line, overlap within the error bars. The vertical dashed line shows the critical value of $\beta$ obtained with the fits~\eq{eq:Binder} of the Polyakov loop susceptibility~\eq{eq:chi2}. We use 100 configurations for all three lattice sizes.}
\label{fig:PL_SU2_Nt2}
\end{figure*}

As the result, the neural network that trained on the value of the $\beta = 4$ deep in the deconfinement region reproduces the Polyakov loop with a perfect agreement with Monte-Carlo data at all other values of the lattice coupling constant including the region of the true deconfinement transition. For the smallest spatial extension, $N_t = 2$, the results are shown in Fig.~\ref{fig:PL_SU2_Nt2}.

The perfect (modulo statistical errors) overlap between the predicted and the original data indicates that the critical value of the coupling constant $\beta_c$ is recovered by the machine-learning algorithm very well. The errors in Fig.~\ref{fig:PL_SU2_Nt2} correspond to the statistical uncertainties inherent to the original Monte Carlo configurations of the gluon fields. At the smallest lattice volume ($N_s = 8$), the statistical errors are naturally larger. We use the same number (100) configurations for all three lattice sizes.

We repeat the same analysis for the lattices with $N_t=4$ in which the critical coupling constant lies in the scaling region of the theory. In this case, to find input data correlations that correspond to the Polyakov Loop, the neural network needs to analyze longer pathways in the gauge groups in order to be able to cover at least one winding of the path along the time direction. Thus, increasing the $N_t$ value requires an additional convolution layer. A combination of two convolution layers allows the machine-learning algorithm to find correlations along four time-links on the lattice. The space of the correlated parameters increases as well. Thus, the dense layer has to contain more neurons to learn the correlations. In the case of $N_t=4$, the dense layer is built of 32 neurons (see Table~\ref{tab:ANN2}). 
\begin{table}[htb]
	\begin{tabular}{|c|c|c|}		
		\hline
	Layer & 	\multicolumn{2}{|c|}{Structure} \\
		\hline
		\multicolumn{3}{|c|}{}\\
		\hline \multirow{2}{*}{InputLayer} & In & ($N_t=4$, $N_s {{\times}} N_s$, $N_s$, $\mbox{Dim} {{\times}} \mbox{U}$) \\
	   \cline{2-3}& Out &($N_t=4$, $N_s {{\times}} N_s$, $N_s$, $\mbox{Dim} {{\times}} \mbox{U}$) \\
		\hline
			\multicolumn{3}{|c|}{}\\
		\hline
\multirow{2}{*}{Conv3D} & In &	($4$, $N_s {{\times}} N_s$, $N_s$, $\mbox{Dim} {{\times}} \mbox{U}$) \\
\cline{2-3} & Out & ($2$, $N_s {{\times}} N_s$, $N_s$, $256$) \\
		\hline
			\multicolumn{3}{|c|}{}\\
		\hline
		\multirow{2}{*}{Conv3D} & In &	($2$, $N_s {{\times}} N_s$, $N_s$, $256$) \\
\cline{2-3} & Out & ($1$, $N_s {{\times}} N_s$, $N_s$, $32$) \\
		\hline
			\multicolumn{3}{|c|}{}\\
		\hline
		\multirow{2}{*}{AveragePooling3D} &In & ($1$, $N_s {{\times}} N_s$, $N_s$, $32$) \\
\cline{2-3}	 & Out & ($1$, $1$, $1$, $32$) \\
		\hline
			\multicolumn{3}{|c|}{}\\
		\hline
		\multirow{2}{*}{Flatten}& In & ($1$, $1$, $1$, $32$) \\
\cline{2-3}		&Out& ($32$)\\
		\hline
			\multicolumn{3}{|c|}{}\\
		\hline
		\multirow{2}{*}{Dense} &In& ($32$) \\
\cline{2-3}	& Out	  & ($1$) \\
		\hline  
\end{tabular}
	\caption{The same as in Table~\ref{tab:ANN2} but for $N_t=4$.}
	\label{tab:ANN2}
\end{table}

\begin{figure}[htb]
\center{\includegraphics[width=0.7\linewidth]{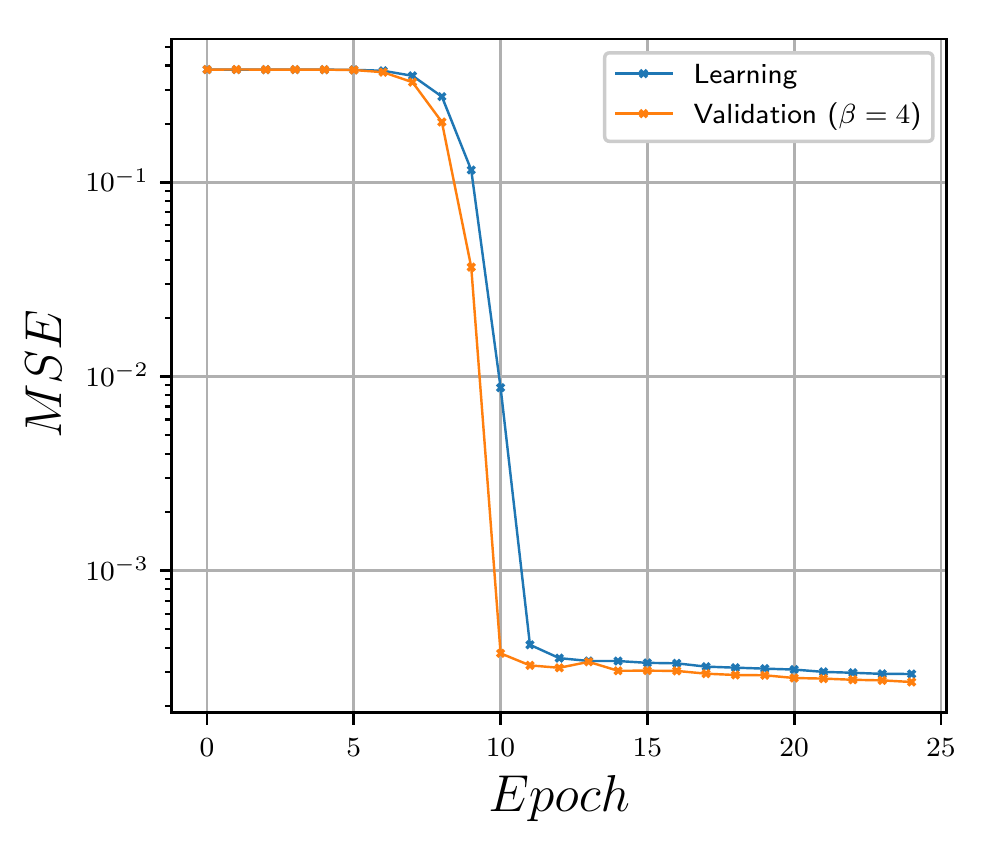}}
\caption{Learning curves for training and validation at the point $\beta = 4$ of the SU(2) gauge theory on $16^3 \times 4$ lattice with the mean squared error (MSE) used as a loss function. The MSE normalized on the value of the order parameter squared, $\langle|L|\rangle^2$, gives qualitatively the same picture.}
\label{fig:Learning:SU2}
\end{figure}

The learning and validation curves for $N_t = 4$ lattice are shown in Fig.~\ref{fig:Learning:SU2}. These are representative examples, qualitatively valid for all studied systems with $N_t = 2, 4$ temporal extensions, $N_s = 8, 16, 32$ spatial sizes, and both SU(2) and SU(3) gauge groups. The learning rate lies in the range [0.001, 0.002] depending on lattice size and theory.
The training with the subsequent validation has been done at the perturbatively deconfining point with $\beta = 4$. 
%In order to check the validity of the results far from the learning point, we made an auxiliary verification at the confining coupling $\beta=0.5$ in the strong coupled region. The latter procedure does not, obviously, influence the training process. The verification is intended to show how well the order parameter, built by the neural network at $\beta = 4$, describes the data at $\beta =0.5$.
Both learning and validation curves of Fig.~\ref{fig:Learning:SU2} show the absence of under- and over-fitting as both curves gradually approach a common plateau at the end of the learning process. 
%The verification curve gives a bit more complex picture: at $\beta = 0.5$ the expectation value of the physical order parameter is very close to zero. The same statement is true for any artificial gauge-dependent polynomial of the gauge fields. At the beginning of the training process, where the unbuilt order parameter is close to an arbitrary combination of the gauge fields, the verification gives small values of the mean squared error. As long as the machine-learning algorithm starts to build a gauge-invariant order parameter, the noise of the gauge group disappears and the MSE of the verification starts to grow thus highlighting the imperfectness of the partially build order parameter. As soon at the learning process gets complete, the MSE of the verification curve acquires an acceptably small and physically justified value. All these properties of the verification curve are clearly seen in Fig.~\ref{fig:Learning:SU2}.

%The most important feature of the successful learning process is that at the final stage of learning, all three curves -- corresponding to training and validation -- overlap at the same plateau value.

The result of the neural network analysis of the $N_t=4$ lattice is presented in Fig.~\ref{fig:PL_SU2_Nt2}. One can clearly see that machine-learning algorithm reproduces the Polyakov loop with a perfect agreement with Monte-Carlo data.

\begin{figure*}[htb]
\begin{tabular}{ccc}
\includegraphics[width=0.3\linewidth]{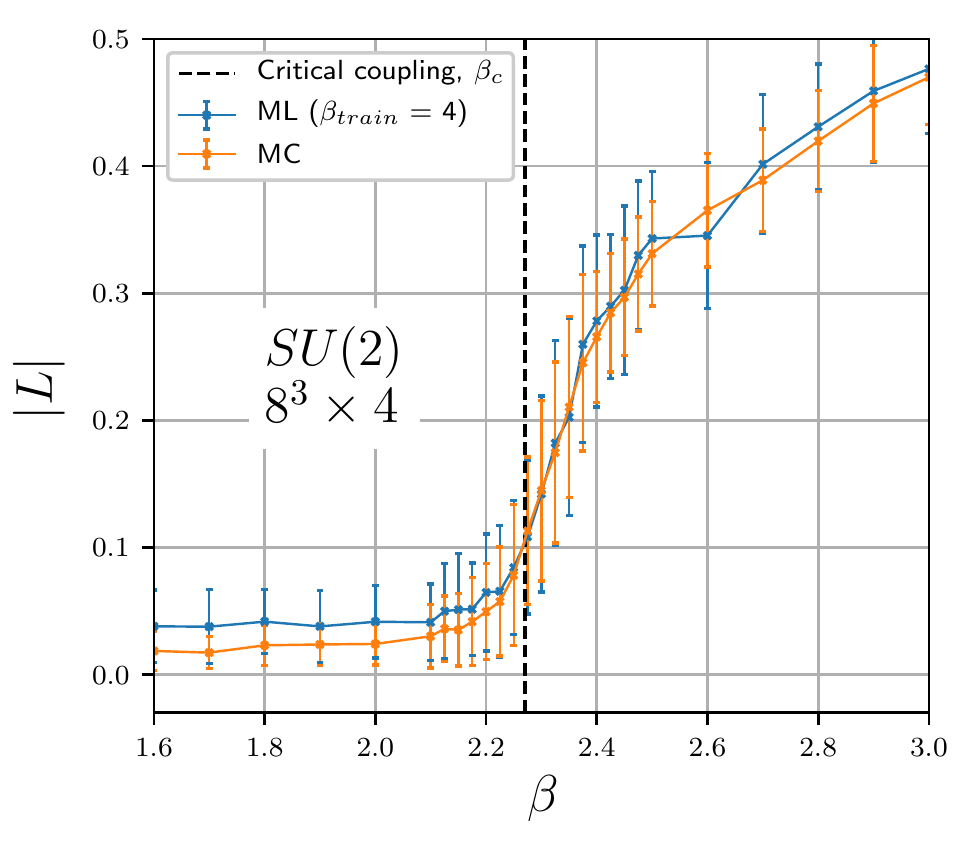} & 
\includegraphics[width=0.3\linewidth]{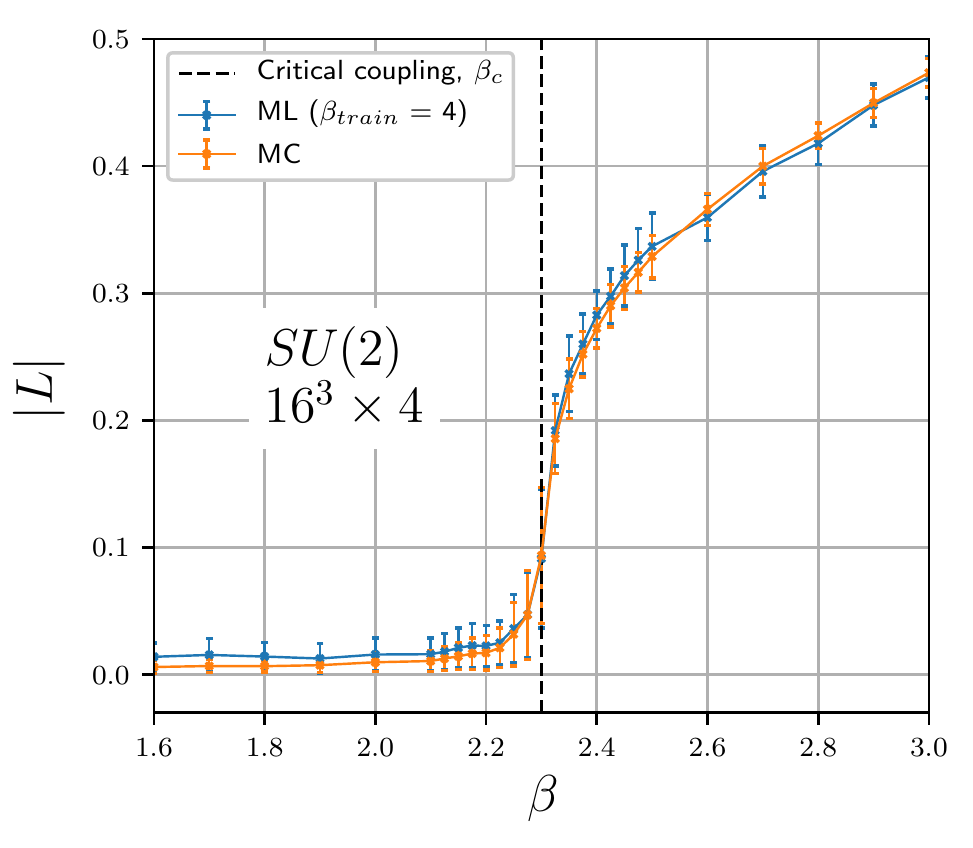} &
\includegraphics[width=0.3\linewidth]{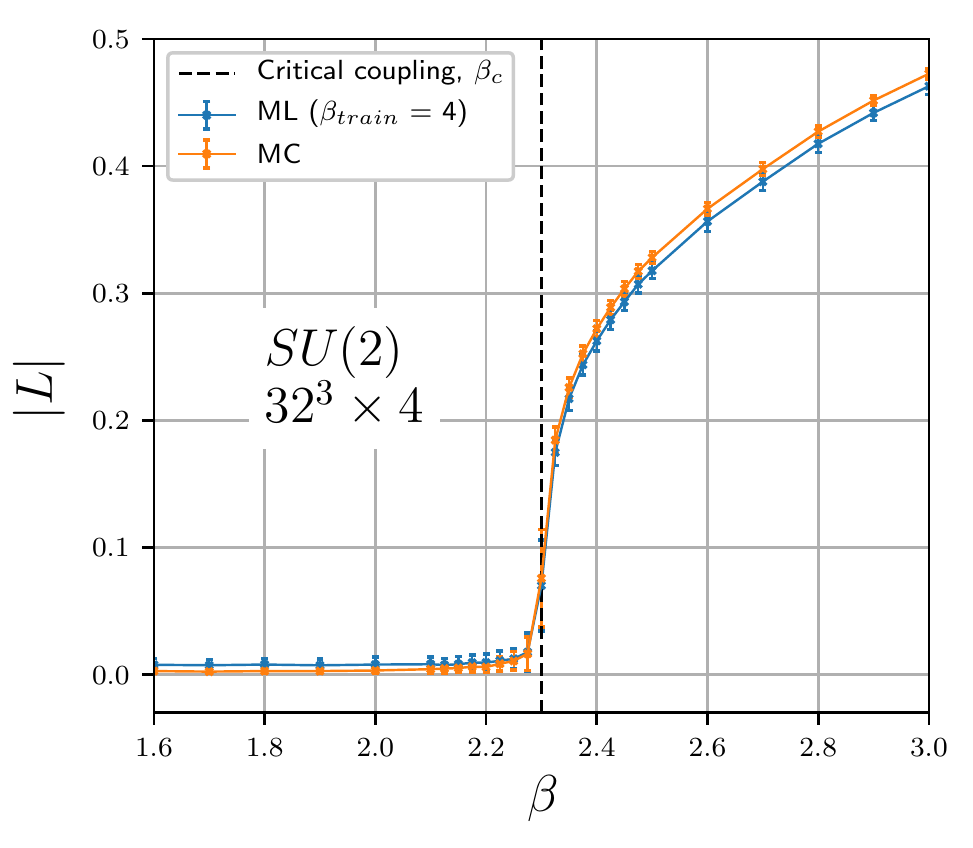}\\[-2mm]
%\hskip 3mm $8^3\times 4$  &
%\hskip 3mm $16^3\times 4$  &
%\hskip 3mm $32^3\times 4$  
\end{tabular}
\caption{The results for the Polyakov loop for SU(2) gauge theory at $N_t = 4$ coming from the Monte Carlo (MC) simulations compared with the prediction of the machine-learning (ML) algorithm. The notations are same as in Fig.~\ref{fig:PL_SU2_Nt2}.}
\label{fig:PL_SU2_Nt4}
\end{figure*}

Our results point to the neural network's ability to find a physically meaningful correlation between the input parameters that correspond to the trace of the SU(2) matrices product along the time direction.  The lattice configurations of the gluon fields generated by the Monte Carlo procedure contain noisy background related to the ultraviolet fluctuations of the gluon fields and random transformations of the SU(2) gauge-symmetry group. The noise ``hides'' the signal of any observables that are not prone to withstand these fluctuations. The ultraviolet fluctuations affect any local observable, while the random gauge transformations hide any non-gauge-invariant quantity in the random noise. 

\begin{figure}[htb]
\center{\includegraphics[width=0.7\linewidth]{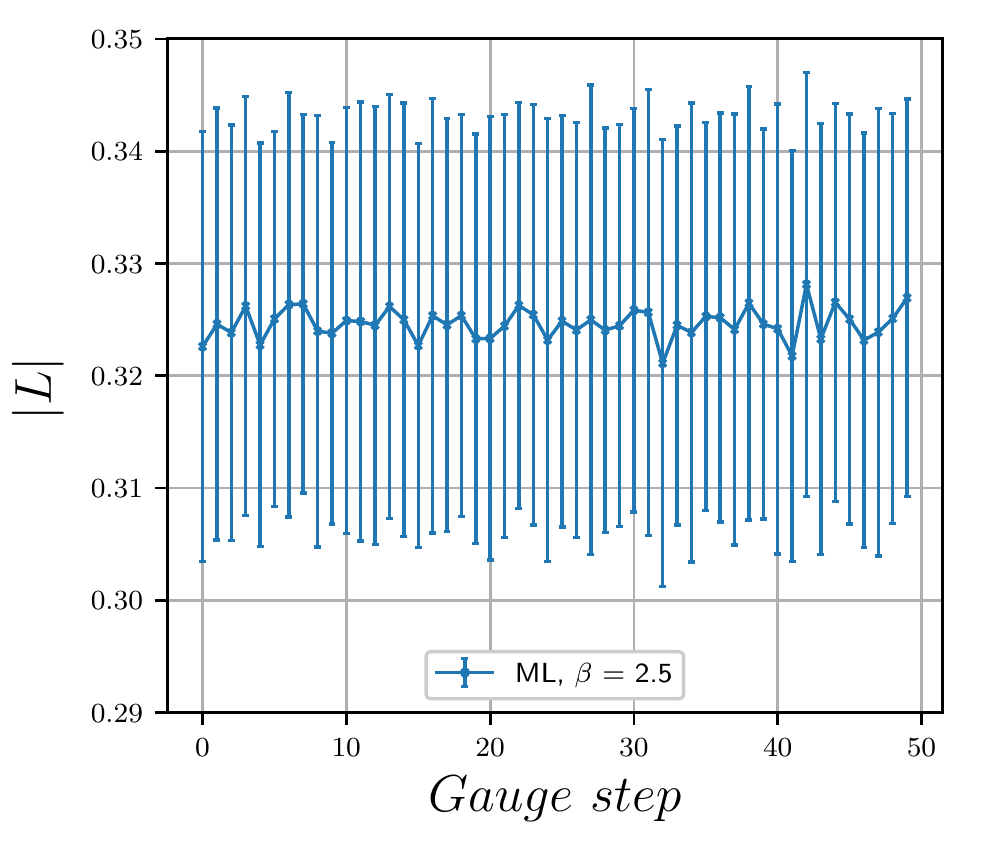}}
\caption{The degree of the gauge dependence in the prediction of the order parameter by the ML algorithm. The predicted order parameter along with the prediction uncertainty vs. the number of the gauge randomization steps of the initial $16^3 \times 4$ gluon configuration at $\beta=2.5$.}
\label{fig:gauge_error:SU2}
\end{figure}

We also check the vulnerability of the ML algorithm for the gauge noise that could theoretically affect the accuracy in the prediction of the Polyakov loop. To this end, we take 100 gluon configurations at the coupling $\beta = 2.5$ for the representative lattice size $16^3\times 4$. We then apply several random gauge transformation to each gluon configuration and subsequently initiate the machine learning algorithm to predict the Polyakov loop using the gauge-randomized gluons as an input. The result, presented in Fig.~\ref{fig:gauge_error:SU2}, shows that the ML algorithm's forecast is a gauge-invariant quantity that does not depend on the strength of the gluonic configuration's randomization in the gauge group's space transformations.

Thus, the neural network selects a non-local and gauge-invariant observable to characterize the phase. This simple observation explains the impressive ability of the machine-learning algorithm to find correlations in the data that correspond to the Polyakov Loop during the learning phase, and subsequently find its values in the full range of the coupling $\beta$ during the prediction phase. 

A correlation between the decision function of the machine-learning algorithm and the Polyakov Loop was pointed out in Ref.~\cite{SU2_ANN}. The correlation was found after the phase classification for the SU(2) theory by polynomial fit of the neural network prediction function. We used a neural network with a 3D convolution layer~(\ref{tab:ANN1}) to analyse the SU(2) group parameters (\ref{input_par}) as independent quantities. Our approach allows us to build and train the neural network that can find the order parameter far outside the range of the lattice coupling values used for the training. As a result, the neural network recovers the order parameter at all physically interesting values of coupling.  

\subsection{SU(3) gauge theory}

\begin{figure*}[!htb]
\begin{tabular}{ccc}
\includegraphics[width=0.3\linewidth]{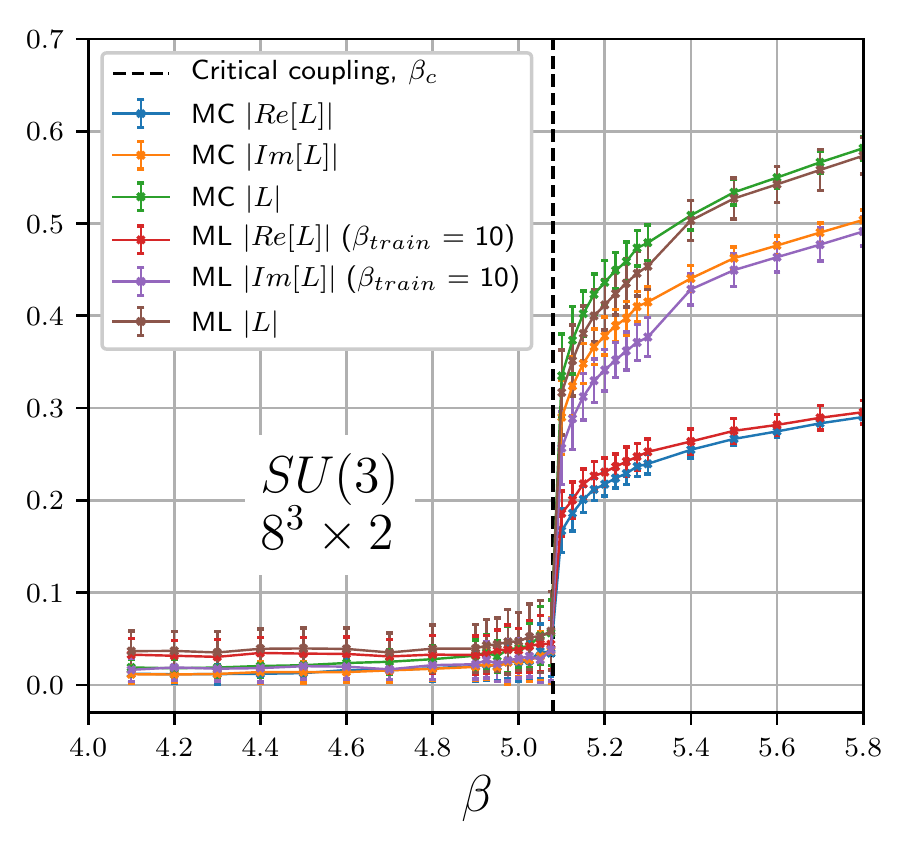} &
\includegraphics[width=0.3\linewidth]{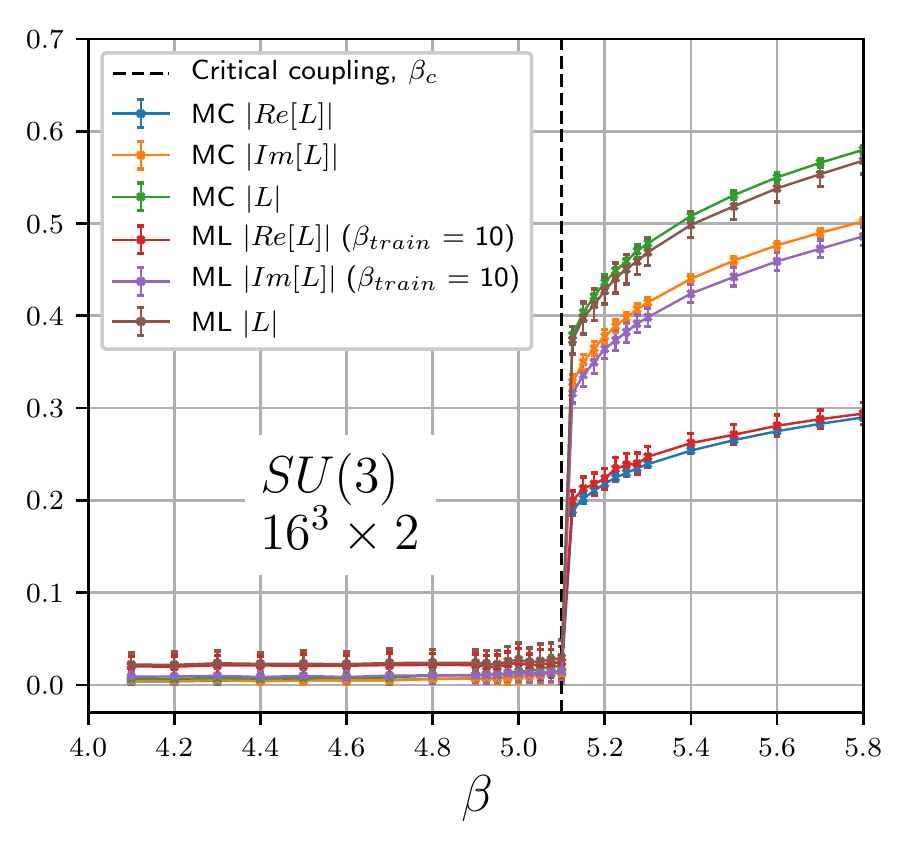} & 
\includegraphics[width=0.3\linewidth]{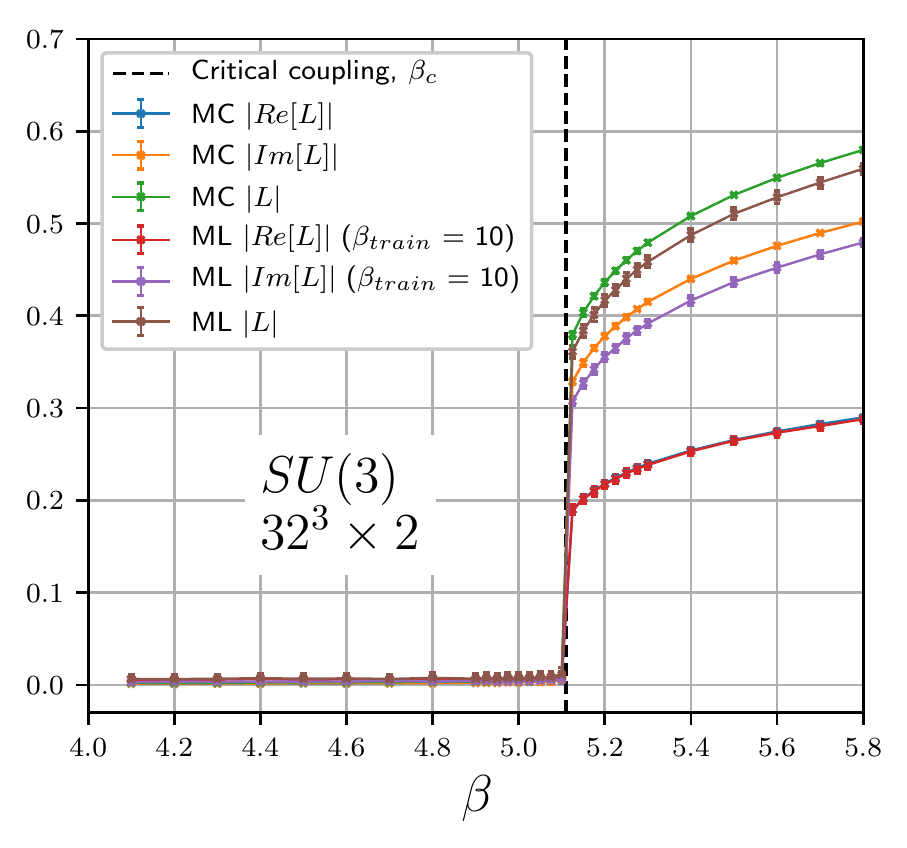} \\[-2mm]
%\hskip 3mm $8^3\times 2$  &
%\hskip 3mm $16^3\times 2$  &
%\hskip 3mm $32^3\times 2$  
\end{tabular}
\caption{The results for the Polyakov loop for SU(3) gauge theory at $N_t = 2$ obtained with the Monte Carlo simulations as compared to the neural network prediction. The absolute value, the real and imaginary parts of the loop are shown. The value of $ML\ |L|$  restored from ML predictions of $|Re[L]|$ and $|Im[L]|$.}
\label{fig:PL_regression4888SU3}
\end{figure*}

In this section we repeat the procedure of the prediction/restoration of the order parameter for the SU(3) configurations. We employ the same architecture of the neural network that has already been used for the SU(2) lattice gauge theory. Contrary to the SU(2) case, we use the full set of 9 complex numbers in the SU(3) case.

In the case of the SU(3) group, the Polyakov loop is a complex number. Therefore, we have to train and predict its value independently both for real and imaginary part of the Polyakov loop. As a training point, we use the lattice coupling $\beta=10$ that corresponds to artificially small lattices which feature the perturbative deconfinement. Similarly to the SU(2) case, we generate 9000 lattice field configurations for the training of the neural network and use only 100 configurations for the prediction. The error bars in the figures reflect the level of statistical fluctuations of the original Monte Carlo configurations.

\begin{figure*}[!htb]
\begin{tabular}{ccc}
\includegraphics[width=0.3\linewidth]{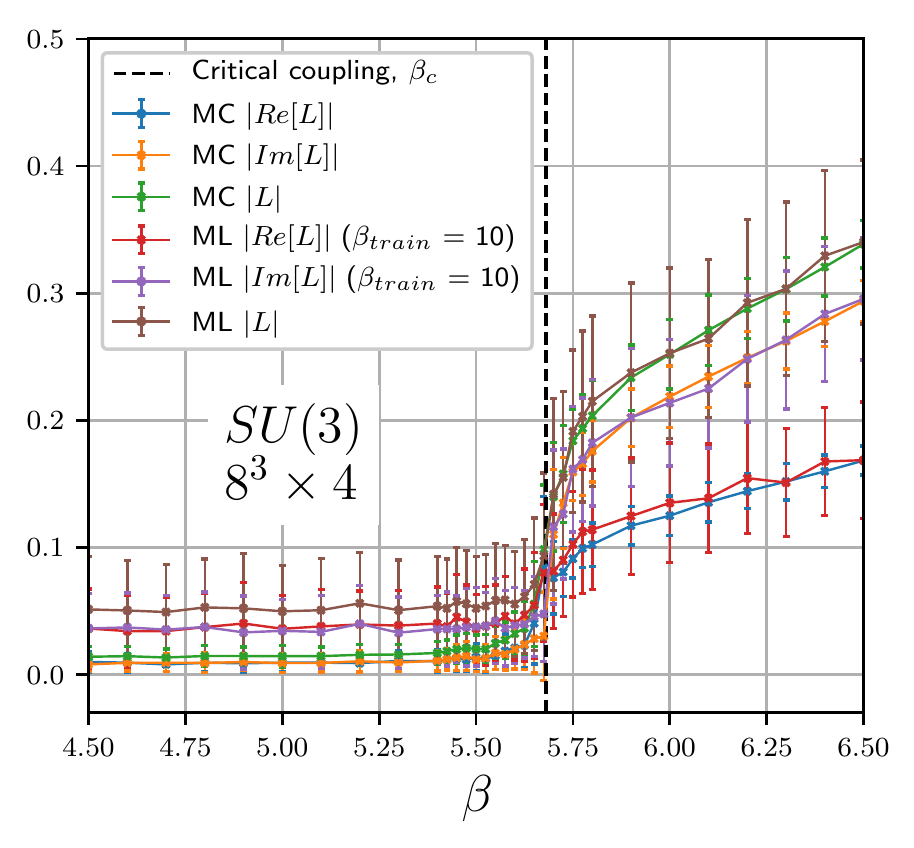} &
\includegraphics[width=0.3\linewidth]{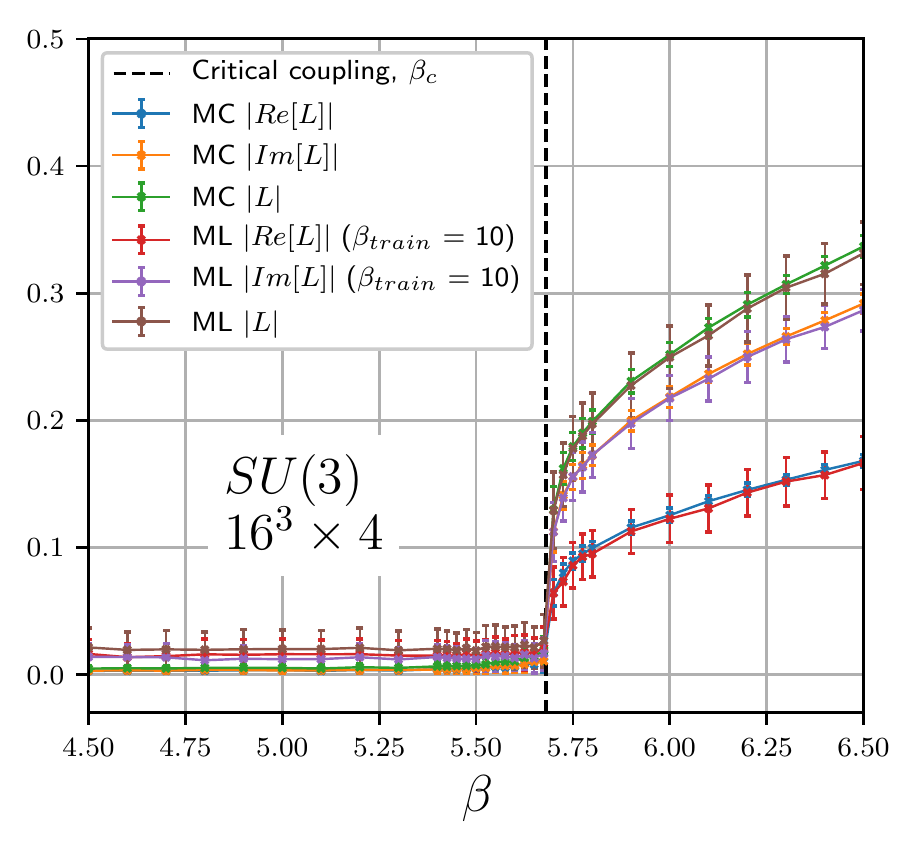} & 
\includegraphics[width=0.3\linewidth]{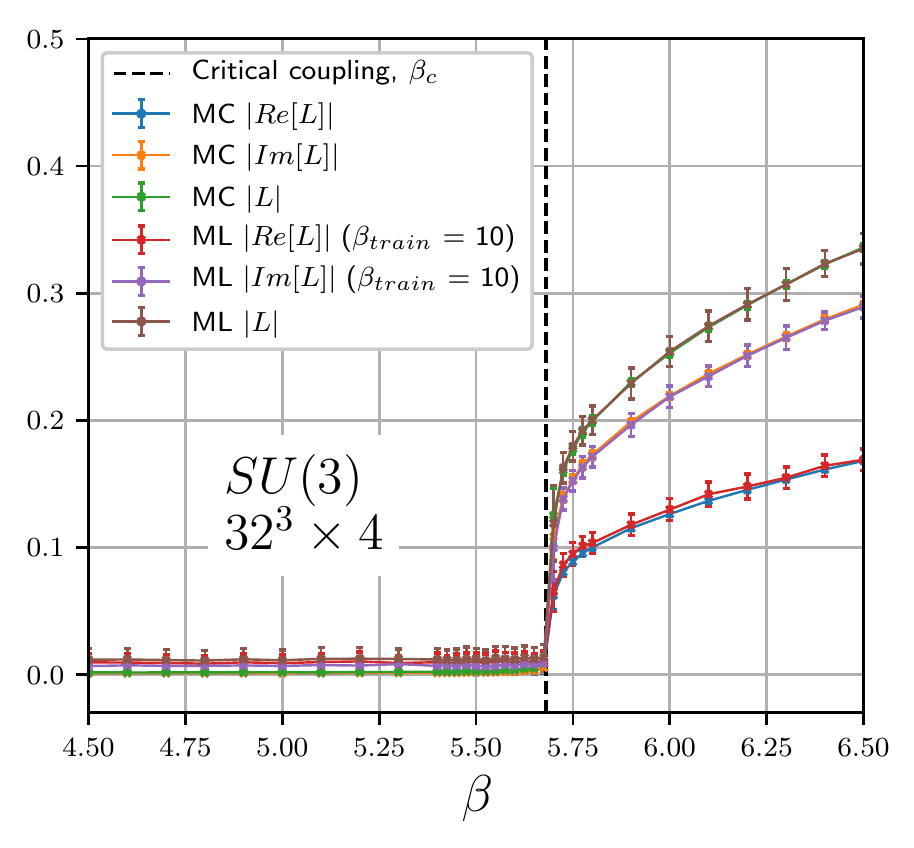} \\[-2mm]
%\hskip 3mm $8^3\times 4$  &
%\hskip 3mm $16^3\times 4$  &
%\hskip 3mm $32^3\times 4$  
\end{tabular}
\caption{The same results as in Fig.~\ref{fig:PL_regression4888SU3}, but for the temporal extension $N_t=4$ of SU(3) gauge theory.}
\label{fig:PL_SU3_Nt4}
\end{figure*}

Repeating the same procedures as we done in the case of SU(2) Yang Mills theory, we obtain the Polyakov loop in a perfect agreement with Monte Carlo simulations of the SU(3) gauge theory. The neural network is able to find the correlations in the lattice data at one (unphysical) point of the lattice coupling and restore the behaviour of this order parameter in the full range of the lattice couplings including the interesting region of the real physical phase transition.

\section{Conclusion}
In our paper, we demonstrated that the neural network may serve as an efficient numerical counterpart of an ``analytical continuation'' of physical observable as a function of  lattice configuration. The machine-learning algorithm allows us to restore a gauge-invariant order parameter in the whole physical region of the parameter space after being trained on lattice configurations at one unphysical point in the lattice parameter space.

We have chosen the training point far away from the physical region at a very weak coupling. This particular choice was deliberately made in the most-possible unphysical way: the training point cannot serve, neither in numerical approaches not in analytical techniques, for any meaningful analysis of the phase structure of the theory because the system experiences a finite-volume deconfinement transition. Therefore, the model resides in the perturbative regime and has no relation to the continuum non-perturbative Yang-Mills theory. 

After the training phase, the neural network was aimed to predict the Polyakov loop as the deconfining order parameter in the SU(2) and SU(3) gauge theories. The machine learning algorithm was able to build a trace of the gauge group matrices product along a closed loop in the time direction. As a result, the neural network trained at one (unphysical) value of the lattice coupling $\beta$ was able to predict the order parameter in the whole region of the $\beta$ values with a good precision. We thus demonstrated that the machine learning techniques may be used as an analytical-type continuation from easily reachable but physically uninteresting regions of the coupling space to the interesting but potentially not accessible regions. This approach may prove to be particularly useful in models, where simulations in a physical region cannot be done to due numerical (computational) constraints provided the unphysical (extreme) points are still available for training.

\acknowledgments

The work of M.N.C, N.V.G, V.A.G, S.D.L, and A.V.M was supported by the grant of the Russian Foundation for Basic Research No.18-02-40121 mega. The numerical simulations of Monte Carlo data were performed at the computing cluster Vostok-1 of Far Eastern Federal University.

\end{document}